\documentstyle[picinpar,cite,axodraw]{prhep97}

\makeatletter
\let\chapter\hid@chapter
\makeatother

\expandafter\ifx\csname mathrm\endcsname\relax\def\mathrm#1{{\rm #1}}\fi

\def\beq{\begin{equation}}
\def\eeq{\end{equation}}
\def\beqar{\begin{eqnarray}}
\def\eeqar{\end{eqnarray}}
\def\barr#1{\begin{array}{#1}}
\def\earr{\end{array}}
\def\bfi{\begin{figure}}
\def\efi{\end{figure}}
\def\btab{\begin{table}}
\def\etab{\end{table}}
\def\bce{\begin{center}}
\def\ece{\end{center}}
\def\nn{\nonumber}

\def\text{\textstyle}

\def\ga{\gamma}
\def\Ga{\Gamma}
\def\de{\delta}

\def\refeq#1{\mbox{(\ref{#1})}}

\def\citere#1{\mbox{Ref.\cite{#1}}}
\def\citeres#1{\mbox{Refs.\cite{#1}}}

\newcommand{\GeV}{\unskip\,\mathrm{GeV}}
\newcommand{\MeV}{\unskip\,\mathrm{MeV}}

\renewcommand{\O}{{\cal O}}

\def\mathswitchr#1{\relax\ifmmode{\mathrm{#1}}\else$\mathrm{#1}$\fi}

\newcommand{\PW}{\mathswitchr W}
\newcommand{\PZ}{\mathswitchr Z}

\newcommand{\Pe}{\mathswitchr e}

\newcommand{\Pepm}{\mathswitchr {e^\pm}}
\newcommand{\Pep}{\mathswitchr {e^+}}
\newcommand{\Pem}{\mathswitchr {e^-}}

\newcommand{\PWpm}{\mathswitchr {W^\pm}}

\def\mathswitch#1{\relax\ifmmode#1\else$#1$\fi}

\newcommand{\MW}{\mathswitch {M_\PW}}

\newcommand{\MZ}{\mathswitch {M_\PZ}}

\newcommand{\Me}{\mathswitch {m_\Pe}}

\def\draftdate{\relax}
\def\mda{\relax}
\def\mua{\relax}
\def\mla{\relax}
\def\draft{
\def\thtystars{******************************}
\def\sixtystars{\thtystars\thtystars}
\typeout{}
\typeout{\sixtystars**}
\typeout{* Draft mode!
         For final version remove \protect\draft\space in source file *}
\typeout{\sixtystars**}
\typeout{}
\def\draftdate{\today}
\def\mua{\marginpar[\boldmath\hfil$\uparrow$]%
                   {\boldmath$\uparrow$\hfil}%
                    \typeout{marginpar: $\uparrow$}\ignorespaces}
\def\mda{\marginpar[\boldmath\hfil$\downarrow$]%
                   {\boldmath$\downarrow$\hfil}%
                    \typeout{marginpar: $\downarrow$}\ignorespaces}
\def\mla{\marginpar[\boldmath\hfil$\rightarrow$]%
                   {\boldmath$\leftarrow $\hfil}%
                    \typeout{marginpar: $\leftrightarrow$}\ignorespaces}
\overfullrule 5pt
\oddsidemargin -15mm
\marginparwidth 29mm
}

\def\stars{\strut\leaders\hbox{*}\hfill\strut}
\def\starline{\hfil\strut\hfil\hbox to \textwidth {\stars}\hfil}

\newcommand{\lsim}
{\;\raisebox{-.3em}{$\stackrel{\displaystyle <}{\sim}$}\;}

\makeatletter
\def\section{\@startsection {section}{1}{\z@}{-15pt plus -4pt minus
-4pt}{9pt plus 4pt minus 4pt}{\Large\bf\boldmath
\pretolerance=10000\relax\rightskip=0pt plus8em}}
\makeatother

\begin{document}

\def\thefootnote{\fnsymbol{footnote}}
\setcounter{footnote}{0}
\null
{\large
\hfill CERN-TH/97-302 \\
\strut\hfill hep-ph/9710542
\vskip 3cm
\begin{center}
{\LARGE\bf
Theoretical aspects of W physics%
\footnote{Contribution to  the proceedings of 
{\it The International Europhysics Conference on
High-Energy Physics}, 19-26 August 1997, Jerusalem, Israel.}
\par} \vskip 4em
{\Large
{\sc Stefan Dittmaier%
}\\[2ex]
{\Large \it Theory Division, CERN\\
CH-1211 Geneva 23, Switzerland}\\[2ex]
}
\par \vskip 1em
\end{center}\par
\vskip .0cm 
\vfill 
{\bf Abstract:} \par 
High-precision predictions for W-production
processes are complicated by the instability of the W~bosons,
requirements of gauge invariance, and the necessity to include radiative
corrections. Salient features and recent progress concerning these issues
are discussed for the process $\Pe\Pe\to\PW\PW\to 4f$.
\par
\vskip 1cm
\noindent
CERN-TH/97-302 \\
October 1997 
\par
\null
\setcounter{page}{0}
\thispagestyle{empty}
}
\def\thefootnote{\arabic{footnote}}
\setcounter{footnote}{0}

\clearpage


\authorrunning{S.\,Dittmaier}
\titlerunning{{\talknumber}: Theoretical aspects of W physics}
 

\def\talknumber{802} 

\title{{\talknumber}: Theoretical aspects of W physics}
\author{Stefan Dittmaier 
(Stefan.Dittmaier@cern.ch)}
\institute{CERN, Theory Division, Switzerland}

\maketitle

\begin{abstract}
High-precision predictions for W-production
processes are complicated by the instability of the W~bosons,
requirements of gauge invariance, and the necessity to include radiative
corrections. Salient features and recent progress concerning these issues
are discussed for the process $\Pe\Pe\to\PW\PW\to 4f$.
\end{abstract}
 
\section{Introduction}

The investigation of the W~boson and its properties at LEP2 
\cite{lep2rep} and possible future linear $\Pep\Pem$ colliders 
\cite{ecfa} is very promising. Together with
the Fermi constant and the LEP1 observables, an 
improvement of the empirical value of the W-boson mass $\MW$ will put 
better indirect constraints on the 
mass of the Standard-Model Higgs boson
and on new-physics parameters. The W-boson mass can
be obtained by inspecting the total W-pair production 
cross-section near threshold, where it is most sensitive to $\MW$, or by
reconstructing the invariant masses of the W~decay products.
W-boson production in $\Pe\Pe$-, $\Pe\gamma$-, and
$\gamma\gamma$-collisions also yields direct information on the vector-boson
self-couplings, which are governed by the gauge symmetry. For low and
intermediate centre-of-mass (CM) energies, useful information can be
obtained by investigating the distributions over the W-production angles.
For higher energies also the total cross-sections become very
sensitive to anomalous couplings.

The described experimental aims require the knowledge of the Standard-Model
predictions for the mentioned observables to a high precision,
e.g.\ for the cross-section of W-pair production at LEP2 to $\sim$0.5\%
\cite{lep2repWcs}.
The instability of the W~bosons, the issue of gauge invariance, and the 
relevance of radiative corrections render this task highly non-trivial. 
In this short presentation these sources of complications and their 
consequences for actual calculations are discussed, and special 
emphasis is laid on recent developments. For definiteness, we consider 
the process $\Pe\Pe\to\PW\PW\to 4f$, which is the most important one for 
W~physics at LEP2. 

\section{Gauge invariance and finite-width effects}

At and beyond a per-cent accuracy, gauge-boson resonances cannot be
treated as on-shell states in lowest-order calculations, since the impact 
of a finite decay width $\Ga_{\mathrm{V}}$ for a gauge boson V of mass 
$M_{\mathrm{V}}$ can be roughly estimated to 
$\Ga_{\mathrm{V}}/M_{\mathrm{V}}$, which is, for instance, $\sim$3\% for the 
W~boson. Therefore, the full set of tree-level diagrams for a given fermionic 
final state has to be taken into account. For $\Pe\Pe\to\PW\PW\to 4f$ this
includes graphs with two resonant W-boson lines (``signal diagrams'') and
graphs with one or no W~resonance (``background diagrams''), leading to
the following structure of the amplitude \cite{lep2repWcs,ae94,be94}:
\looseness -1
\beqar
\label{eq:Mstruc}
{\cal M} =
\underbrace{\frac{R_{+-}(k_+^2,k_-^2)}{(k_+^2-\MW^2)(k_-^2-\MW^2)}
        }_{\mbox{doubly-resonant}}
+\underbrace{\frac{R_{+}(k_+^2,k_-^2)}{k_+^2-\MW^2}
            +\frac{R_{-}(k_+^2,k_-^2)}{k_-^2-\MW^2}
        }_{\mbox{singly-resonant}}
+\underbrace{N(k_+^2,k_-^2)}_{\mbox{non-resonant}}\!.
\nn\\[-1em] \\[-2.3em] \nn
\eeqar
Gauge invariance implies that ${\cal M}$ is independent of
the gauge fixing used for calculating Feynman graphs (gauge-parameter
independence), and that gauge cancellations between different
contributions to ${\cal M}$ take place.
These gauge cancellations are ruled by Ward identities.
For a physical description of the W~resonances, the finite W~decay width
has to be introduced in the resonance poles.
However, since only the sum in
\refeq{eq:Mstruc}, but not the single contributions to ${\cal M}$, 
possesses the gauge-invariance properties, the simple replacement
\beq
\left[k_\pm^2-\MW^2\right]^{-1} \quad\to\quad
\left[k_\pm^2-\MW^2+i\MW\Ga_\PW(k_\pm^2)\right]^{-1}
\eeq
in general violates gauge invariance. 

Although such gauge-invariance-breaking terms are formally suppressed by 
a factor $\Gamma_{\mathrm{V}}/M_{\mathrm{V}}$ \cite{ae93}, 
they can completely destroy 
the consistency of predictions if they disturb gauge cancellations
\cite{ku95,imflscheme,flscheme}. 
Gauge cancellations can occur if a current $\bar u(p_1)\gamma^\mu u(p_2)$
that is associated to a pair of external fermions
becomes proportional to the momentum $k$ of the attached gauge boson:
\\[.3em]
\hspace*{4em}
{\unitlength1pt 
\begin{picture}(240,60)(0,0)
\ArrowLine(15,55)(40,30)
\ArrowLine(40,30)(15, 5)
\Photon(40,30)(90,30){2}{5}
\GCirc(40,30){8}{.5}
\Vertex(90,30){1.2}
\ArrowLine(90,30)(115,55)
\ArrowLine(115, 5)(90,30)
\LongArrow(60,40)(70,40)
\put(92,49){$p_1$}
\put(92,10){$p_2$}
\put(63,49){$k$}
\put(63,14){$V$}
\Vertex(50,20){0.6}
\Vertex(47,18){0.6}
\Vertex(42,17){0.6}
\Vertex(50,40){0.6}
\Vertex(47,42){0.6}
\Vertex(42,43){0.6}
\put(140,30)
{$\sim\quad\displaystyle\frac{1}{k^2-M_{\mathrm{V}}^2}\;k^\mu T^V_\mu.$}
\end{picture}
} 
\\
$T^V_\mu$ represents the set of subgraphs hidden in the blob.
The cancellations in $k^\mu T^V_\mu$ are governed by the Ward identities
\beq
\label{eq:WIs}
k^\mu T^\ga_\mu     = 0, \qquad
k^\mu T^Z_\mu       = i\MZ T^\chi, \qquad
k^\mu T^{W^\pm}_\mu = \pm\MW T^{\phi^\pm}.
\eeq
The first one expresses electromagnetic current conservation and is
relevant, e.g., for forward scattering of $\Pepm$ ($k\to 0$). 
The others imply the Goldstone-boson equivalence theorem, which relates 
the amplitudes for high-energetic longitudinal W and Z bosons 
($k^0\gg M_{\mathrm{V}}$) to the ones for their respective would-be 
Goldstone bosons $\phi$ and $\chi$.

Among the proposed methods (see \citeres{lep2repWcs,flscheme} and
references therein) to introduce finite widths for W and Z~bosons in
tree-level amplitudes, the field-theoretically most convincing one is
provided by the ``fermion-loop scheme''. This scheme
goes beyond a pure tree-level calculation by including and consistently
Dyson-summing all closed fermion loops in $\O(\alpha)$. This procedure
introduces the running tree-level width in gauge-boson
propagators via the imaginary parts of the fermion loops. 
The Ward identities \refeq{eq:WIs} are not violated, since the
fermion-loop (as well as the tree-level) contributions to vertex functions 
obey the simple linear (also called ``naive'') Ward identities that are 
related to the original gauge invariance rather than to the more involved 
BRS invariance of the quantized theory.
Owing to the linearity of the crucial Ward identities for the vertex
functions, the fermion-loop scheme works both with the full fermion
loops and with the restriction to their imaginary parts.
The full fermion-loop scheme has been worked out for 
$\Pe\Pe\to\PW\PW\to 4f$ in \citere{flscheme}, where
applications are discussed as well. Simplified versions of the
scheme have been introduced in \citere{imflscheme}.

The fermion-loop scheme is not applicable in the presence of resonant
particles that do not exclusively decay into fermions. For such
particles, parts of the decay width are contained in bosonic
corrections. The Dyson summation of fermionic {\it and}
bosonic ${\cal O}(\alpha)$ corrections leads to inconsistencies in the
usual field-theoretical approach, i.e.\ the Ward identities
\refeq{eq:WIs} are broken in general.
This is due to the fact that the
bosonic ${\cal O}(\alpha)$ contributions to vertex functions do not obey 
the ``naive Ward identities''. The problem is circumvented by employing
the background-field formalism \cite{de95}, in which these naive
identities are valid. This implies \cite{de96} that a consistent Dyson
summation of fermionic and bosonic corrections to any order in $\alpha$
does not disturb the Ward identities \refeq{eq:WIs}.
Therefore, the background-field approach provides a natural generalization
of the fermion-loop scheme.
We recall that any resummation formalism goes beyond a strict
order-by-order calculation and necessarily involves ambiguities
in relative order $\alpha^n$ if not all $n$-loop diagrams are included. 
This kind of scheme dependence, which in particular concerns gauge 
dependences, is only resolved by successively calculating the missing
orders.

Note that the consistent resummation of all 
${\cal O}(\alpha)$ loop corrections does not automatically lead to 
${\cal O}(\alpha)$ precision in the predictions if resonances are involved.
The imaginary parts of one-loop self-energies generate only
tree-level decay widths so that directly on resonance one order in
$\alpha$ is lost. To obtain also full ${\cal O}(\alpha)$
precision in these cases, the imaginary parts of the two-loop
self-energies are required. However, how and whether this two-loop
contribution can be included in a practical way without violating the
Ward identities \refeq{eq:WIs} is still an open problem. 
Taking the imaginary parts of all two-loop contributions solves the 
problem in principle at least for the background-field approach, but 
this is certainly impractical.

Fortunately, the full off-shell calculation for the process 
$\Pe\Pe\to\PW\PW\to 4f$ in ${\cal O}(\alpha)$ is not needed for
most applications. Sufficiently above the W-pair threshold 
a good approximation should be obtained by taking into account
only the doubly-resonant part of the amplitude \refeq{eq:Mstruc}, 
leading to an error of the order of $\alpha\Ga_\PW/(\pi\MW)\lsim 0.1\%$. 
In such a ``pole scheme'' calculation \cite{ae94,st91} the numerator 
$R_{+-}(k_+^2,k_-^2)$ has to be replaced by the gauge-independent residue 
$R_{+-}(\MW^2,\MW^2)$. 
The structure of this approach, which is in fact
non-trivial and has not been completely carried out so far, is
described below.

\pagebreak

\section{Electroweak radiative corrections}

Present-day Monte Carlo generators
for off-shell W-pair production (see e.g.\ \citere{lep2repWevgen})
include only universal electroweak ${\cal O}(\alpha)$ corrections%
\footnote{The QCD corrections for hadronic final states are discussed 
in \citere{qcdcorr}.}
such as the running of the electromagnetic coupling, $\alpha(q^2)$, 
leading corrections entering via the $\rho$-parameter,
the Coulomb singularity \cite{coul}, which is important near threshold,
and mass-singular logarithms 
$\alpha\ln(\Me^2/Q^2)$ from initial-state radiation. 
In leading order, the scale $Q^2$ is not determined and has to be set to
a typical scale for the process; for the following we take $Q^2=s$.
Since the full ${\cal O}(\alpha)$ correction is not known for off-shell
W~pairs, the size of the neglected ${\cal O}(\alpha)$ contributions is
estimated by inspecting on-shell W-pair production, for which the exact 
${\cal O}(\alpha)$ correction and the leading contributions were
presented in \citeres{rceeww} and \cite{bo92}, respectively.
These ${\cal O}(\alpha)$ corrections have already been implemented in an 
event generator for on-shell W~pairs \cite{ja97}.
The following table shows the difference between an ``improved
Born approximation'' $\de_{\mathrm{IBA}}$, which is based on the 
above-mentioned universal corrections, and the corresponding full
${\cal O}(\alpha)$ correction $\de$ to the Born cross-section integrated over
the W-production angle $\theta$ for some CM energies $\sqrt{s}$.
\\[.5em]
\hspace*{4em}
\begin{tabular}{|c||c||c|c|c|c|c|c|}
\hline
$\theta$ range & $\sqrt{s}/\GeV$ &
161 & 175 & 200 & 500 & 1000 & 2000 \\
\hline\hline
$0^\circ$$<$$\theta$$<$$180^\circ$ & 
$(\de_{\mathrm{IBA}}-\de)/\%$ 
& 1.5 & 1.3 & 1.5 & 3.7 & 6.0 & 9.3 \\
\cline{1-1} \cline{3-8}
$10^\circ$$<$$\theta$$<$$170^\circ$ &&
1.5 & 1.3 & 1.5 & 4.7 & 11 & 22 \\
\hline
\end{tabular}
\\[.5em]
Here the corrections $\de_{\mathrm{IBA}}$ and $\de$ include only soft-photon
emission. For
more details and results we refer to \citeres{lep2repWcs,crad96}.
The quantity $\de_{\mathrm{IBA}}-\de$ corresponds to the neglected
non-leading corrections and amounts to $\sim$1--2\% for LEP2 energies,
but to $\sim$10--20\% in the TeV range. Thus, in view of the aimed
$0.5$\% level of accuracy for LEP2 and all the more for energies of
future linear colliders, the inclusion of non-leading corrections is
indispensable. 
The large contributions in $\de_{\mathrm{IBA}}-\de$ at high energies are
due to terms such as $\alpha\ln^2(s/\MW^2)$, which arise from vertex 
and box corrections and can be read off from the high-energy 
expansion \cite{be93} of the virtual and soft-photonic ${\cal O}(\alpha)$ 
corrections.
\looseness -1

As explained above, a reasonable starting point for incorporating 
${\cal O}(\alpha)$ corrections beyond universal effects is provided by a
double-pole approximation. Doubly-resonant corrections to 
$\Pe\Pe\to\PW\PW\to 4f$ can be classified into two types: factorizable
and non-factorizable corrections \cite{lep2repWcs,ae94,be94}.
The former are those that correspond either to W-pair production 
\cite{rceeww} or to W~decay \cite{rcwdecay}. 
Since these corrections were extensively discussed in
the literature, we focus on the non-factorizable corrections.
They are furnished by diagrams in which the production
subprocess and/or the decay subprocesses are not independent. Among such
corrections, doubly-resonant contributions only arise if a ``soft'' photon
of energy $E_\gamma \lsim \Ga_\PW$
is exchanged between the subprocesses. 

In \citere{fa94} it was shown that the non-factorizable corrections 
vanish if the invariant masses of both W~bosons are integrated over. 
Thus, these corrections do not influence pure angular distributions,
which are of particular importance for the analysis of gauge-boson
couplings. For exclusive quantities the non-factorizable corrections are 
non-vanishing and have been calculated in \citeres{me96,be97,de97}%
\footnote{The recent evaluations \cite{be97,de97} are in complete mutual
agreement, but confirm the analytical results of \citere{me96} only for
the special final state $f\bar f f'\bar f'$.}.
It turns out \cite{de97} that the correction factor to the differential 
Born cross-section is non-universal in the sense that it depends on the
parametrization of phase space. The calculations \cite{me96,be97,de97}
have been carried out using the invariant masses $M_\pm$ of the \PWpm~bosons, 
which are identified with the invariant masses of the respective final-state
fermion pairs, as independent variables. 
Since all effects from the initial $\Pep\Pem$ state
cancel, the resulting correction factor does not depend on the
W-production angle and is also applicable to processes such as
$\gamma\gamma\to\PW\PW\to 4f$. 
Figure~\ref{fig:nonfactnum} shows that non-factorizable
corrections to a single invariant-mass distribution are of the order of
$\sim$1\% for LEP2 energies, shifting the maximum of the distribution by an
amount of 1--2$\MeV$, which is small with respect to the experimental
uncertainty at LEP2 \cite{lep2repWmass}. For higher energies the 
non-factorizable correction is more and more suppressed.
\begin{figure}[h]
\centerline{
\setlength{\unitlength}{1cm}
\begin{picture}(8,5.8)
\put(0,0){\includegraphics{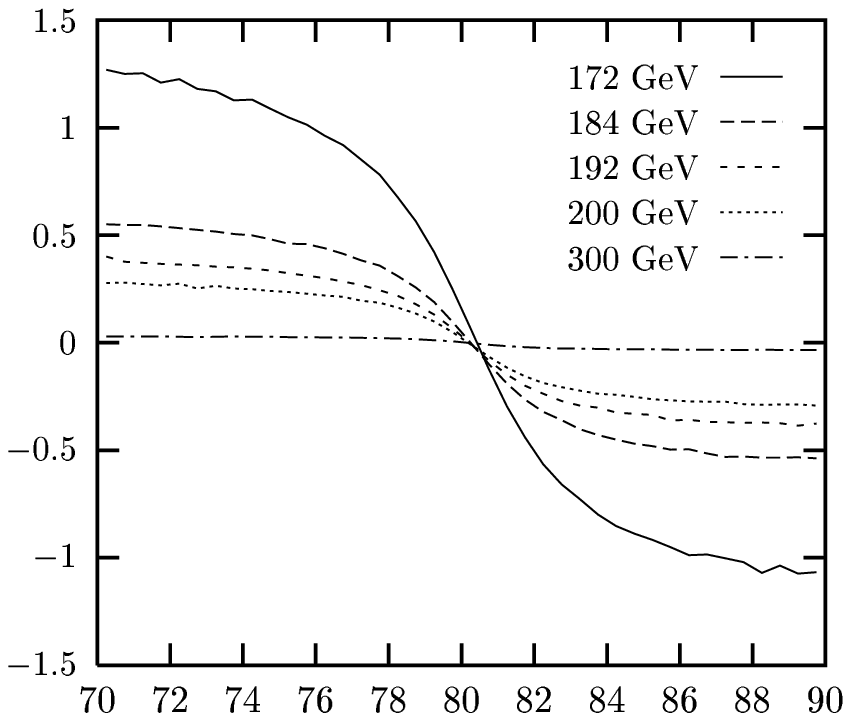}}
\put(0.0,4.0){\makebox(1,1)[c]{$\de_{\mathrm{nf}}/\%$}}
\put(4.0,-0.7){\makebox(1,1)[cc]{{$M_+/\GeV$}}}
\end{picture}
}
\caption{Relative non-factorizable corrections to the invariant-mass
distribution $\protect\mathrm{d}\sigma/\protect\mathrm{d} M_+$ 
in $\Pep\Pem\to\mu^+\nu_\mu\tau^-\bar\nu_\tau$
for some CM energies (plot taken from \citere{de97}).}
\label{fig:nonfactnum}
\efi

%

\end{document}